\documentclass[aps,pra,twocolumn,nobalancelastpage,superscriptaddress]{revtex4-1}
\pdfoutput=1

\usepackage{graphicx,color,amsmath,amsfonts,amssymb,mathtools,dsfont,verbatim}
\usepackage{epstopdf}

\newcommand{\diag}[1]{\mathrm{diag} \begin{pmatrix} #1 \end{pmatrix}}

\DeclarePairedDelimiter\bra{\langle}{\rvert}
\DeclarePairedDelimiter\ket{\lvert}{\rangle}

\DeclarePairedDelimiterX\braket[2]{\langle}{\rangle}{#1 \delimsize\vert #2}

\usepackage{braket}

\begin{document}

\title{$\mathbb{Z}_N$ symmetry breaking in Projected Entangled Pair State models}

\author{Manuel Rispler}
\affiliation{Max-Planck-Institute of Quantum Optics, Hans-Kopfermann-Str.\ 1, 85748 Garching, Germany}
\affiliation{JARA Institute for Quantum Information, RWTH Aachen University, 52056 Aachen, Germany}

\author{Kasper Duivenvoorden}
\affiliation{JARA Institute for Quantum Information, RWTH Aachen University, 52056 Aachen, Germany}

\author{Norbert Schuch}
\affiliation{Max-Planck-Institute of Quantum Optics, Hans-Kopfermann-Str.\ 1, 85748 Garching, Germany}

\begin{abstract}

We consider Projected Entangled Pair State (PEPS) models with a global
$\mathbb Z_N$ symmetry, which are constructed from $\mathbb
Z_N$-symmetric tensors and are thus $\mathbb Z_N$-invariant wavefunctions,
and study the occurence of long-range order and symmetry breaking in these
systems. First, we show that long-range order in those models is
accompanied by a degeneracy in the so-called transfer operator of the
system.  We subsequently use this degeneracy to determine the nature of
the symmetry broken states, i.e., those stable under arbitrary perturbations, and
provide a succinct characterization in terms of the fixed points of the
transfer operator (i.e.\ the different boundary conditions) in the
individual symmetry sectors. We verify our findings numerically through
the study of a $\mathbb Z_3$-symmetric model, and show that the
entanglement Hamiltonian derived from the symmetry broken states is
quasi-local (unlike the one derived from the symmetric state), reinforcing
the locality of the entanglement Hamiltonian for gapped phases.

 \end{abstract}

\maketitle

\section{Introduction}

Spontaneous symmetry breaking is a prime example of the emergence of
global order from local interactions in quantum systems at zero
temperature.  The formation of macroscopic domains in which a single
ordered state is selected from a set of energetically equivalent states is
witnessed by the onset of long-range order, i.e., non-decaying two-point
correlations.  In any finite system, the ground space of a symmetric
Hamiltonian is an irreducible representation, and therefore unique for
any Abelian symmetry.  Long range order implies the existence of low lying
excited states~\cite{Griffiths,komotasaki} for which the gap closes in the
thermodynamic limit. The physical ground states are then those which are
stable under general (symmetry breaking) perturbations of the Hamiltonian;
they  turn out to be hybridizations of the symmetric and low lying excited
states and therefore break the symmetry of the system. 

Projected Entangled Pair States (PEPS) form a framework for modelling the
low-energy states of interacting quantum systems.  The central object here
is a local tensor which is being used to build up global wavefunctions
locally, based on their entanglement structure. PEPS thus form the right
ansatz to approximate the low-energy physics of systems governed by
local interactions~\cite{hastings:locally,molnar:thermal-peps}, making
them a powerful tool for the variational simulation of interacting
many-body systems~\cite{verstraete:2D-dmrg,orus:tn-review}. At the same
time, PEPS form a versatile analytical framework: Since every PEPS is the
exact ground state of an associated parent
Hamiltonian~\cite{ParentHamiltonian} which inherits all symmetries from
the tensor, they can be used to construct solvable models where the
desired physical structure is built directly into the tensor. A
particularly appealing feature of PEPS models is that they allow to
explicitly identify the degrees of freedom associated to the entanglement
spectrum and the edge physics of the system. Thereby, they clarify the nature
of the one-dimensional system underlying both edge physics and
entanglement properties, and allow to explicitly determine
the one-dimensional entanglement
Hamiltonian~\cite{cirac:peps-boundaries,yang:peps-edgetheories}.

Despite their costruction from local tensors, PEPS models can naturally
describe systems with emergent global order, such as systems with
topological entanglement, or systems which exhibit long-range order and
thus spontaneous symmetry breaking~\cite{verstraete:comp-power-of-peps}.
Yet, in the scenario of spontaneous symmetry breaking the PEPS tensor, which
encodes the local physics of the system, will clearly be invariant under
the respective symmetry, and thus will also be the global PEPS wavefunction.
This is, while the system exhibits
long-range order, the wavefunction does not actually break the symmetry,
which is reflected in unphysical cat-like states in the entanglement
spectrum of the system.  Thus, the question arises how to understand the
occurence of spontaneous symmetry breaking in PEPS models with long-range
order, and in particular how to obtain the symmtry broken wavefunctions and the
corresponding edge states and entanglement spectra and Hamiltonians.

In this paper, we study the occurrence of symmetry breaking for PEPS models
with an abelian $\mathbb Z_N$ symmetry. Specifically, we address two
questions: First, we show how long-range order in a PEPS model is
accompanied by a degeneracy in the so-called transfer operator, and
second, we use this degeneracy to determine the structure of the symmetry
broken states, i.e., those ground states which are stable under perturbations.
We then apply our results to study the entanglement
Hamiltonian, where we observe that the symmetry
broken states allow to restore the locality of the entanglement
Hamiltonian in the symmetry broken phase.

More specifically, we start by considering a system with long-range order,
which generally implies the presence of symmetry breaking.  In PEPS, the
so-called transfer operator (describing a one-dimensional slice of the
system) mediates all correlation functions.  We prove that the presence of
long-range order implies an approximate degeneracy in the spectrum of the
transfer operator, labelled by symmetry sectors, which becomes exact in
the thermodynamic limit. The different fixed points of the transfer
operator correspond to different states in the ground space manifold of
the system.  We then consider the behavior of the transfer operator under
physical perturbations of the model (i.e., those corresponding to
perturbations of its Hamiltonian). Using the algebraic structure of the
fixed point space, we are able to succinctly characterize the stable fixed
points, and we find that there is a unique set of stable fixed points,
given by the Fourier transform of the fixed points in the individual irrep
sectors.  These stable fixed points provide the boundary conditions which
yield the  symmetry broken states. At the same time, any fixed point also
provides direct access to the entanglement spectrum of the
systems~\cite{cirac:peps-boundaries}.  We use this to derive the
entanglement Hamiltonian both for the symmetric and the symmetry broken
ground states, and find that the locality of the entanglement Hamiltonian
is restored by choosing the symmetry broken states.  This reinforces the
perspective that the entanglement Hamiltonian is local for any gapped
phase~\cite{cirac:peps-boundaries,schuch:topo-top}.

We have initiated the study of PEPS with symmetry breaking and long-range
order in Ref.~\cite{rispler:lro-and-ssb-in-peps}, where we have considered
the special case of a broken $\mathbb Z_2$ symmetry, related it to the
transfer operator spectrum, and determined the stable fixed points.  In
the present work, we  generalize this to the case of $\mathbb Z_N$
symmetries, to which end we in particular establish entirely different
proof techniques in order to characterize the stable fixed points, based
on the algebra structure of the fixed point space of the transfer
operator.  The present approach also goes beyond
Ref.~\cite{rispler:lro-and-ssb-in-peps} in that it no longer requires
Hermiticity of the transfer operator.

The paper is structured as follows: In Sec.~\ref{sec:peps-models}, we
introduce the necessary tools to study PEPS models.  In
Sec.~\ref{sec:longrangeorder} we review and extend the arguments of
Ref.~\cite{rispler:lro-and-ssb-in-peps} to show that long range order
implies a degeneracy in the transfer operator. In Sec.~\ref{sec:proof} we
show which of these fixed points model stable environments. These stable
fixed points can be obtained from symmetric fixed points by a Fourier
transform.  In Sec.~\ref{sec:numerics} we perform a numerical study of a
family of $\mathbb{Z}_3$-symmetric PEPS models related to the $\mathbb
Z_3$ Potts model to verify the stability of the earlier defined stable
fixed points, and demonstrate that the entanglement Hamiltonian arising
from stable fixed points is quasi-local.

\section{\label{sec:peps-models}Projected Entangled Pair State models}
Let us introduce PEPS. We will, w.l.o.g., work on a square lattice of size $N_v\times N_h.$ The model is defined by the five-index tensor $A_{\alpha\beta\gamma\delta}^i$ with the \emph{physical} index $i = 0\dots d-1$, where $d$ is the physical dimension of each site and the \emph{auxiliary} indices $\alpha,\beta,\gamma,\delta = 0\dots D-1$ with the so called \emph{bond dimension} $D$. This tensor could be site dependent, but to ease notation, we stick with a single site independent tensor and in doing so only consider translational invariant models. The wave function can be constructed by putting one tensor on each site of the lattice and contracting all \emph{auxiliary} indices. One could either decide to model periodic boundary conditions (contract indices on opposite edges) or model open boundary conditions by an extra boundary tensor. The remaining, \emph{physical} indices constitute the coefficients of the wavefunction as $\ket{\Psi} = \sum c_{i_1\dots i_{(N_vN_h)}}\ket{i_1\dots i_{(N_vN_h)}}$.

We will assume the tensor $A_{\alpha\beta\gamma\delta}^i$ to be symmetric under a $\mathbb{Z}_N$ symmetry generated by unitary $S$ and $s$ in the following manner:
\begin{equation}\label{symmetry-relation}
\sum_j s_{ij}A^j_{\alpha\beta\gamma\delta} =
\sum_{\alpha'\beta'\gamma'\delta'}
A^{i}_{\alpha'\beta'\gamma'\delta'}S_{\alpha\alpha'}S_{\beta\beta'}S_{\gamma'\gamma}^\dag
S_{\delta'\delta}^\dag \ .
\end{equation}
It is straightforward to see that the global wave function is invariant as
$s^{\otimes {(N_v N_h)}} \ket{\psi} = \ket{\psi}$ in case of periodic
boundary conditions. The PEPS tensors $A$ not only give rise to a wave
function, but also to a parent Hamiltonian which has this wave function
as its ground state, and moreover commutes with a global action of the
symmetry $s$. In order to define the parent Hamiltonian, let $\mathcal{A}$
be a linear map from the auxiliary space to the physical space,
$\mathcal{A}: (\mathbb{C}^{D})^{\otimes 4}\rightarrow \mathbb{C}^{d}$, related to the
PEPS as $\mathcal{A} =\sum
A_{\alpha\beta\gamma\delta}^i\ket{i}\bra{\alpha\beta\gamma\delta}$.
Similarly, for any region $R$ one can construct the linear map
$\mathcal{A}_R$ from the boundary auxiliary space of $R$ to the bulk
physical space of $R$ by taking $|R|$ copies of $\mathcal{A}$ and
contracting the inner indices. The parent Hamiltonian is given by $H =
\sum_Rh_R$ where $h_R$ acts as a the projector on the orthogonal
complement of the image of $\mathcal{A}_R$. The sum runs for example over
all $R$ forming a 2 by 2 patch. From the symmetry of $A$, all maps
$\mathcal{A}_R$ are also symmetric and hence also their image, showing
that the parent Hamiltonian is also symmetric.

\begin{figure}[!t]
  \includegraphics[width=0.15\textwidth]{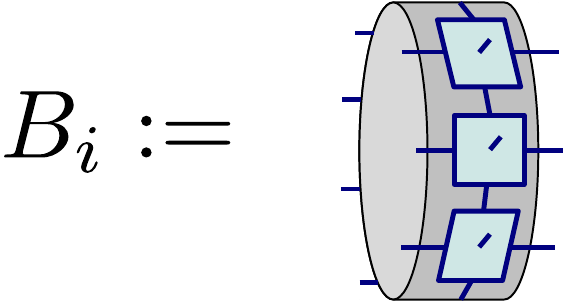}
  \caption{\label{fig:def_Bi}
Definition of the $B_i$ tensor, where all left virtual indices, all physical indices and all right virtual indices are each viewed as the respective composite indices of an MPS tensor.}
\end{figure}

\begin{figure}[!t]
  \includegraphics[width=0.3\textwidth]{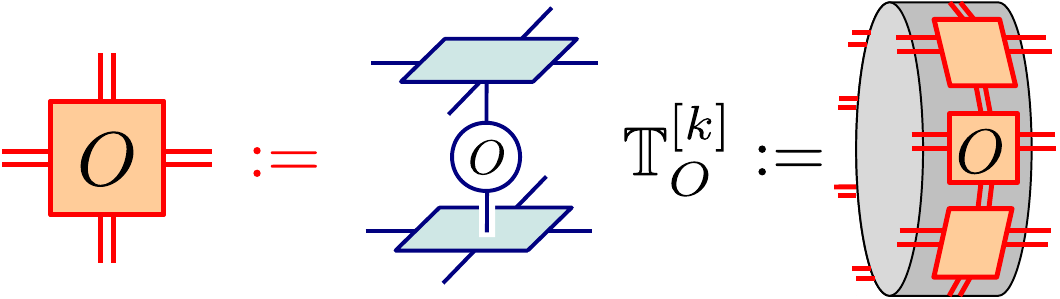}
  \caption{\label{fig:TO}
Definition of the dressed transfer operator $\mathbb{T}^{[k]}$, the
operator acts on physical level on site $[k]$ (site label omitted in
figure).}
\end{figure}

For the rest of this
article we will assume periodic boundary conditions in the vertical ($y$)
direction, thus either toric or cylindrical topology. By blocking the
tensors corresponding to each vertical slice of the lattice into one
composite tensor $B_i$, Fig.~\ref{fig:def_Bi}, the state formally becomes
a matrix product state (MPS).  The transfer operator for such states is a
completely positive map defined as $\mathbb{T}(\rho) := \sum_i B^i \rho
(B^i)^\dagger$. Similar to the MPS case, it can be used to calculate wave
function overlaps and expectation values. For example let $O$ and $O'$ be
two operators acting on sites $i = (i_x,i_y)$ and $j = (j_x,j_y)$, with
$i_x<j_x$. Let $\mathbb{T}^{[k]}_O$ be the mixed transfer matrix obtained
by inserting an operator $O$ between the local tensors $A$ and $A^\dagger$
corresponding to the $k-th$ site: $\mathbb{T}^{[k]}_O := \sum_{ij} B^j
\rho (B^i)^\dagger O^{[k]}_{ij}$, see Fig.~\ref{fig:TO}.

Then we have that 
\begin{align}\label{eq:expvalues}
 \braket{O_iO'_j} = \frac{\text{Tr}\left[ l^\dagger \mathbb{T}^{d_1}\mathbb{T}^{[i_y]}_{O}\mathbb{T}^{d_2} \mathbb{T}^{[j_y]}_{O'}\mathbb{T}^{d_3}(r)\right]}{\text{Tr}\left[ l^\dagger \mathbb{T}^{N_h}(r)\right]} \ ,
\end{align}
with $d_1 = i_x-1$, $d_2=j_x-i_x-1$ and $d_3 = N_h-j_x$ and $l$ and $r$
some boundary tensors. From this expression we see that, just as in the
one dimensional case, all correlations between observables on different
horizontal sites are controlled by the spectral properties of the transfer
operator.  A new feature in two dimensions is that this transfer operator
now has itself a one dimensional structure, which as we will see later
allows for criticality and phase transitions. These features do not arise
in the one dimensional MPS case, where correlations are guaranteed to
decay exponentially since $\mathbb{T}$ is independent of system size.

The symmetry of the tensor $A$ carries over to the Kraus operators $B_i$
of the transfer operator, $\sum_i u_{ij} B_j = U B_iU^\dagger$, and hence to
the transfer operator: 
\begin{equation}\label{eq:symT}
U\mathbb{T}(\rho)U^\dagger = \mathbb{T} (U\rho U^\dagger) \ ,
\end{equation}
where $U = S^{\otimes N_v}$ and $u=s^{\otimes N_v}$.  The eigenvectors of
$\mathbb{T}$ can thus be labeled by their symmetry. For each symmetry sector $\alpha\in0\dots N-1$, let $\lambda_\alpha$ be the largest eigenvalue, with corresponding eigenvector $r_\alpha$. That is $\mathbb{T}(r_\alpha) = \lambda_\alpha r_\alpha$ and 
\begin{align}\label{eq:sym_fp}
 Ur_\alpha U^\dagger= \omega^\alpha r_\alpha \ ,
\end{align}
with $\omega = \exp(\frac{2\pi i}{N})$. Clearly $r_\alpha$ is not positive
for any $\alpha\neq0$, it is not even Hermitian unless
$\omega^\alpha\in\mathbb{R}$ since then $r_\alpha^\dagger$ transforms according
to the irrep $(\omega^{\alpha})^*\ne \omega^\alpha$ (star denotes complex
conjugation). From complete positivity of $\mathbb{T}$, its eigenvector
corresponding to the largest
eigenvalue should be positive and hence $|\lambda_0|\geq|\lambda_\alpha|$.
We will assume that the eigenvalue $\lambda_\alpha$ is non-degenerate for each
symmetry sector and that $|\lambda_0|>|\lambda_\alpha|$ unless an onset of
an order parameter enforces them to be equal. Non-degeneracy in the
trivial symmetry sector implies, by positivity of $\mathbb{T}$ and $r_0$,
that $\lambda_0>0$. The PEPS tensors can be rescaled as $A \rightarrow
\lambda_0^{\frac{1}{2N_v}}A$ to ensure that $\lambda_0 = 1$, making $r_0$ a fixed
point of the transfer operator. We will refer to any eigenvector having
eigenvalue 1 as fixed point, and their span as the fixed point space, of
$\mathbb{T}$.

From Eq.~\eqref{eq:expvalues} it is also clear that order parameters
$\braket{Z}$ (with $uZ=\gamma\, Zu$, $\gamma\neq 1$) vanish by symmetry if the boundary
tensors $l$ and $r$ are symmetric ([$l,S] = [r,S] = 0$). Also, for large
$d_3$, $\mathbb{T}^{d_3}(r)$ will converge to an eigenvector of
$\mathbb{T}$ with largest eigenvalue and having overlap with $r$.
Similarly for $[\mathbb{T}^*]^{d_1}(l)$, where $\mathbb{T}^*$ is the dual
transfer operator: $\mathbb{T}^*(\rho) := \sum_i (B^i)^\dagger \rho B^i$.
This suggest that, in the thermodynamic limit, symmetry breaking occurs if
there exist a largest eigenvector of the transfer operator which is not
symmetric. In the following section we will elaborate on this statement.

\section{\label{sec:longrangeorder} Long range order in the transfer operator}

Given a local Hamiltonian $H_0$ with a symmetry $[H,u]=0$, there are two
ways to define symmetry breaking: The first is a non-vanishing spontaneous
magnetization
\begin{align}
m:=\lim_{B\rightarrow0}
\lim_{\Lambda\rightarrow\infty}\frac{1}{N}\braket{O}_{B,\Lambda}\ ,
\end{align}
and the other one a non-zero long-range order
\begin{equation}
\sigma :=
\lim_{\Lambda\rightarrow\infty}\frac{1}{N}\sqrt{\braket{O^\dagger O}_{0,\Lambda}}  \ ,
\end{equation}
for some suitably chosen magnetization operator $O =
\sum_{i\in\Lambda}Z_i$ (with local operators $Z_i$).  Here,
$\Lambda$ refers to the set of all sites, and
$\langle\:\cdot\:\rangle_{B,\Lambda}$ denotes the expectation value in the
ground state of the Hamiltonian with a symmetry breaking field,
$H_\Lambda(B) = H_0+B \sum_{i\in\Lambda} Z_i$. It has been shown in a
number of cases that $m\ge \sigma$, i.e., long-range order implies a
non-zero spontaneous magnetization~\cite{KHvdL,komotasaki}.  It is for
this reason that we consider PEPS wavefunctions with long-range order
(which we will use interchangably with symmetry breaking in the
following); our goal will be on the one hand to understand the conditions
under which long-range order occurs, and on the other hand to identify the
wavefunctions describing the corresponding symmetry broken states (i.e.,
those obtained as ground states of $H_\Lambda(B)$ in the limit
$B\rightarrow0$).

Specifically, in the case of a $\mathbb{Z}_N$ symmetry considered in this
work, long-range order will denote a non-zero $\sigma$ for some $Z$ 
obeying
\begin{align}
\label{graded-commutativity} u^\dagger Zu = \omega^\alpha Z \ ,
\end{align} 
for some $\alpha = 0\dots N-1$ and again  $\omega =
\exp(\frac{2\pi i}{N})$. The $\alpha = 1$ case we will refer to as full
symmetry breaking since $Z$ does not commute with any symmetry operation
$u^n$. On the other hand, if $\mathrm{gcd}(\alpha,N)>1$, then the symmetry is only
partially broken since $Z$ commutes with $u^{N/\mathrm{gcd}(\alpha,N)}$.
The advantage of long range order as apposed to a non-zero order parameter
for the detection of symmetry breaking is that a symmetric state can have
long range order whereas any non-symmetric operator has zero expectation
value with respect to a symmetric state. Evaluating $m$ will require a
symmetry breaking field and hence a non symmetric PEPS, which would not
help us in understanding how a symmetric PEPS could describe ordered
phases. Evaluating $\sigma$ can be done using a symmetric PEPS: we will
even do so on a closed manifold  (periodic boundary conditions) such that
the full state is symmetric, as discussed in the previous section.
 
\begin{figure}[!t]
  \includegraphics[width=0.2\textwidth]{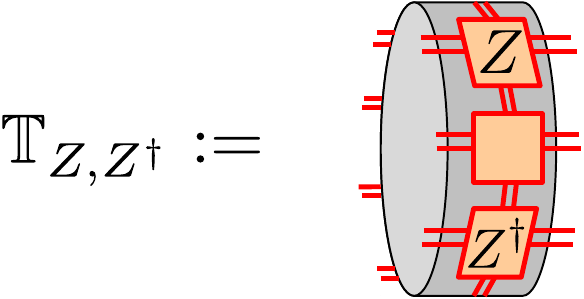}
  \caption{\label{fig:TZZ}The transfer operator $\mathbb{T}_{Z,\bar{Z}^\dagger}^{[i,j]}$ dressed on two physical sites, where the full operator is given by all permutations}
\end{figure}

Let us now turn towards PEPS. For a PEPS $\ket{\Psi}$, 
we have that $\sigma^2 =  \lim_{N_h, N_v
\rightarrow \infty }\sigma^2_{N_v,N_h}$ where 
\begin{equation}
\sigma^2_{N_v,N_h} = \frac{1}{N_h^2 N_v^2} \sum_{ij}\frac{\bra{\Psi} Z_i
Z^\dagger_j \ket{\Psi}}{\braket{\Psi|\Psi}}\ . 
\end{equation} 
In what follows, we will normalize $Z_i$ such that its
operator norm $\|Z_i\|_\mathrm{op}\le 1$.
We will
decompose the above sum over $i$ and $j$ into two parts: either $i_y\neq
j_y$ or $i_y= j_y$. For the first part, define $\mathbb{T}_Z :=
\frac{1}{N_v}\sum_{k=1}^{N_v} \mathbb{T}_Z^{[k]}$ and for the second part
define  $\mathbb{T}_{Z,Z^\dagger} := \frac{1}{N^2_v}\sum_{i,j=1}^{N_v}
\mathbb{T}_{Z,Z^\dagger}^{[i,j]}$  where
$\mathbb{T}_{Z,\bar{Z}^\dagger}^{[i,j]}$ is the transfer operator obtained
by inserting an operator $Z$ and $Z^\dagger$ at sites $i$ and $j$
respectively, see Fig.~\ref{fig:TZZ}.

Define $C_1$ and $C_2$ as 
\begin{align}\label{eq:C1}
 C_1 &= \sum_{p=0}^{N_h-2}\text{Tr}\left[\mathbb{T}_{Z^\dagger} \mathbb{T}^{p} \mathbb{T}_{Z}\mathbb{T}^{N_h-p-2}\right]\ ,\\
 C_2 &= \text{Tr}\left[\mathbb{T}_{Z,Z^\dagger}\mathbb{T}^{N_h-1}\right]\ .
\end{align}
This allows us to write $\sigma^2_{N_v,N_h} = \frac{1}{N_h}\frac{C_1+C_2}{\text{Tr}\left[ \mathbb{T}^{N_h}\right]}$. The factors of $N_v$ are taken care of by the definition of $\mathbb{T}_{Z,Z^\dagger} $ and $\mathbb{T}_Z$ and a factor of $N_h$ is taken care of by using translation invariance in the horizontal direction. The contribution of $C_2$ converges to zero in the large $N_h$ limit. It corresponds to taking the sum over $N_v^2N_h$ expectation values and dividing by $N_v^2N_h^2$. The term of interest is $C_1$.

From the scaling of $\frac{C_1}{N_h\text{Tr}\left[ \mathbb{T}^{N_h}\right]}\propto \mathcal{O}(1)$ we aim to show that the gap between $|\lambda_\alpha|$ and the largest eigenvalue $\lambda_0$ decreases with increasing $N_h$. A first step in the proof is that for large $N_h$, $\mathbb{T}^{N_h}\rightarrow |r_0)(l_0|$, where $r_0$ is the eigenvector of $\mathbb{T}$ corresponding  to the eigenvalue $\lambda_0$ and $l_0$ the corresponding left eigenvector (ie. eigenvector of $\mathbb{T}^*$). We use round brackets to emphasis that, although $r_0$ and $l_0$ are eigenvectors, they are also operators. The dressed transfer operator $\mathbb{T}_{Z}$ maps $r_0$ into the symmetry sector $\alpha=1$ due to
\begin{align}\nonumber
 U \mathbb{T}_{Z}(r_0) U^\dagger &=  \mathbb{T}_{u^\dagger Z u}(Ur_0U^\dagger) \\
 &=\omega \mathbb{T}_{Z }(r_0)\ .
\end{align}
More explicitly, if $UlU^\dagger = \omega^\beta l$, then
$\text{Tr}\left[l^\dagger\mathbb{T}_{Z}(r_0)\right] \neq 0$ only if $\beta
= \alpha$. The main idea is that the factor $\mathbb{T}^p$ will give rise
to an exponential suppression, the leading term being proportional to
$|\lambda_\alpha|^p$. So $|\lambda_\alpha|<1$ will result in zero long
range order. However, as $N_v$ increases, the dimension of the space on
which $\mathbb{T}$ acts also increases exponentially as $D^{2N_v}$, where
$D$ is the bond dimension of the PEPS tensor $A$. Without any other
assumptions, large Jordan blocks could prevent exponential suppression. As
an example, consider a map $T$ in Jordan form with a single Jordan block
of size $D$ and corresponding eigenvalue $\lambda<1$. Then for
$v=(0,\dots,0,1)^T$,
$\|T^pv\|^2=\sum_{q=0}^{\mathrm{min}(p,D)}\lambda^{2(p-q)}\binom{p}{q}^2$.
For large $p$
this sum scales as $\lambda^{2p}p^{2D}$, which for constant $D$ is eventually
exponentially suppressed but only at a length scale $p \propto D$. Hence,
due to the exponentially increasing dimension of $\mathbb{T}$,
correlations are only suppressed only over a length $D^{2N_v}$ leading to
a scaling of the long range order as $\sigma^2 \propto
\frac{1}{N_h}D^{2N_h}$, even in the case that $|\lambda_\alpha|<1$. It is
for this reason that we need more assumptions on $\mathbb{T}$.

We will assume for the rest of the section that the transfer operator
$\mathbb{T}$ is normal, $\mathbb T\mathbb T^*=\mathbb T^*\mathbb T$.  This
is in particular the case if $\mathbb T$ is Hermitian, which for example
can follow from Hermiticity of its Kraus operators $(B^i)^\dagger = B^i$
which physically is related to a combination of time reversal (complex
conjugation) and reflection along the $y$-axis (transposition) symmetry.  

Let us now return to Eq.~\eqref{eq:C1}. In
\cite{rispler:lro-and-ssb-in-peps} it was shown that if the largest
eigenvalue $\lambda_0=1$ of $\mathbb{T}$ is non degenerate, then for any
bounded operator $O$ the following holds: 
\begin{align}\nonumber
\lim_{N_h\rightarrow\infty}&\sum_{p=0}^{N_h-2}\frac{\text{Tr}\left[\mathbb{T}_{O^\dagger}
\mathbb{T}^{p} \mathbb{T}_{O}\mathbb{T}^{N_h-p-2}\right]}{\text{Tr}\left[
\mathbb{T}^{N_h}\right]} = \\ 
\label{ourresult}
&2\sum_{p=0}^{\infty}\text{Tr}\left[\mathbb{T}_{O^\dagger} \mathbb{T}^{p}
\mathbb{T}_{O}|r_0)(l_0|\right]\ . 
\end{align} 
The factor of two arises from first splitting the sum into two parts, one
for which $p> N_h/2$ and one for which $p< N_h/2$. Both sums are identical
(if $N_h$ is odd) up to the position of the dagger, which can be
swapped using ket-bra-hermiticity which exchanges $\mathbb
T_O\leftrightarrow\mathbb T_O^\dagger$, while leaving the other terms
unchanged, and thus $\sum_{p=0}^{N_h-2} \rightarrow
2\sum_{p=0}^{N_h/2-1}$.  (The original proof does not carry the dagger,
but this can be easily adapted.)

Using Eq.~\eqref{ourresult}, we now have that
\begin{align}
\lim_{N_h \rightarrow \infty }\!\!N_h\sigma^2_{N_v,N_h}\!
&= 
    \sum_{p=0}^{\infty}\text{Tr}\left[\mathbb{T}_{Z^\dagger} \mathbb{T}^{p} \mathbb{T}_{Z}|r_0)(l_0|\right]
\nonumber
\\
&= 
    \sum_{p=0}^{\infty}(l_0|\mathbb{T}_{Z^\dagger} 
    (P_\alpha\mathbb{T}P_\alpha)^{p} \mathbb{T}_{Z}|r_0)
\nonumber
\\
&\stackrel{(a)}{=} 
    (l_0|\mathbb{T}_{Z^\dagger} 
    (1-P_\alpha\mathbb{T}P_\alpha)^{-1} \mathbb{T}_{Z}|r_0)
\nonumber
\\
&\stackrel{(b)}{\le} 
    \|(l_0|\mathbb{T}_{Z^\dagger}\|_2
    \|(1\!-\!P_\alpha\mathbb{T}P_\alpha)^{-1}\|_\mathrm{op}
\|\mathbb{T}_{Z}|r_0)\|_2
\nonumber
\\
&\le \frac{1}{1-|\lambda_\alpha|}\ ,
\nonumber
\end{align}
where $P_\alpha$ is the projector onto the irrep sector $\alpha$.
It is in $(a)$ and $(b)$ that we have used normality of $\mathbb T$, which
implies that $\|P_\alpha\mathbb TP_\alpha\|_\mathrm{op}<1$ such that the
Neumann series converges, and $\|\mathbb T_Z|r_0)\|_2^2 = (r_0|\mathbb
T_Z\mathbb T_Z|r_0)\le 1$, as $\|Z_i\|_\mathrm{op}\le 1$ and the left and
right eigenvectors coincide.

We now have 
\begin{equation}
\label{eq:onelimit-twolimits}
\sigma^2 = \!\! 
    \lim_{N_h,N_v \rightarrow \infty }\!\!\sigma^2_{N_v,N_h}
\le
    \lim_{ N_v \rightarrow \infty }\frac{1}{N_v} \lim_{N_h \rightarrow \infty }N_h\sigma^2_{N_v,N_h}\ ,
\end{equation}
where the inequality can be shown by coupling the l.h.s.\ limit such that
$N_h$ grows sufficiently faster than $N_v$, based on the formal
definition of the limit~\footnote{%
Let $\tau_{N_v} = \lim_{N_h} N_h \sigma^2_{N_h,N_v}$, and let
$S:=\lim_{N_v}\tau_{N_v}$.  Then, for any $\epsilon>0$, 
\begin{align*}
&\exists N_v^0 \ \forall N_v\ge N_v^0\,:\ 
    \Big| \tfrac{1}{N_v} \tau_{N_v} - S \Big| < \frac{\epsilon}{2}
\mbox{\quad and}
\\
&\exists N_h^0(N_v) \ \forall N_h\ge N_h^0(N_v):\:
    \frac{1}{N_v}\Big| N_h \sigma_{N_h,N_v}^2 - \tau_{N_v} \Big| <
\frac{\epsilon}{2}\ . \hspace*{-2em}
\end{align*}
Thus, 
\[
\left|\frac{N_h}{N_v}\sigma_{N_h,N_v}^2-S\right|\le \epsilon \ ,
\]
and finally
\[
\sigma^2=\lim_{N_h,N_v\to\infty} \sigma_{N_h,N_v}^2 \le \frac{N_h}{N_v}\sigma_{N_h,N_v}^2=S \ ,
\]
as long as we couple the limits such that both $N_h\ge N_h^0(N_v)$ and $N_h\ge N_v$. (If
$S=\infty$, the inequality \eqref{eq:onelimit-twolimits} holds trivially.)
Let us note that for normal $\mathbb T$, the convergence in
Eq.~\eqref{ourresult}, Ref.~\cite{rispler:lro-and-ssb-in-peps}, yields a
scaling $N_h^0(N_v)\propto N_v/(1-|\lambda_\alpha(N_v)|)$, such that
a non-zero $\sigma^2$ for all isotropically coupled limits
$N_h/N_v=\mathrm{const.}$ is sufficient to
infer that $|\lambda_\alpha|\rightarrow 1$ as $N_v\to\infty$.
}. It thus follows that
if $\sigma^2>0$, for sufficiently large $N_v$ it must hold that
\begin{align}
0 < \frac{1}{N_v} \ \frac{1}{1-|\lambda_\alpha|}\ .
\end{align}
Thus non zero long range order of an order parameter $Z$ obeying
Eq.~\eqref{graded-commutativity} for some $\alpha$ implies that
$|\lambda_\alpha| \ge 1-\mathcal{O}(1/N_v)$. 

At this point we have not said
anything about the phase of $\lambda_\alpha$. It is known that
peripheral spectrum (eigenvalues of modulus 1) consists of roots of unity
\cite[Proposition 3.3]{fannes1992} and that any eigenvalue of the form
$e^{2\pi i/p}$ corresponds to a $p$ periodic state. The degeneracy of the
fixed point would then relate to a breaking of translation symmetry, as
well as the global symmetry $s$. An example is the antiferomagnetic phase.
We can remove such a phase by blocking $p$ sites, i.e.\  consider
$\mathbb{T}^p$, yielding that $\lambda_\alpha \ge 1-O(1/N_v)$.

\section{\label{sec:proof}Boundary of ground states}
Ground states in an ordered phase are not only eigenstates of a symmetric Hamiltonian, but also eigenstates of any perturbed Hamiltonian. Similarly, we will discuss in this section, which fixed points of the transfer operator are also fixed points of any perturbed transfer operator. In more detail, let $H = \sum_Rh_R$ be the parent Hamiltonian of some PEPS defined by the tensors $A^{[i]}$. Consider the perturbation in which each local term is conjugated by an operator of the form $\Lambda_R = \Lambda^{\otimes|R|}$ where $\Lambda$ is close to unity. The perturbed Hamiltonian 
\begin{align}\label{eq:Hpert}
H_\Lambda = \sum_R(\Lambda^{-1})^\dagger_R h_R\Lambda^{-1}_R\ ,
\end{align}
is clearly positive and annihilates the perturbed PEPS
$\ket{\Psi_\Lambda}$ constructed from perturbed tensors $A^{i}_\Lambda  :=
\sum_j\Lambda_{ij} A^{j}$. Hence $\ket{\Psi_\Lambda}$ is the ground state
of $H_\Lambda$.  We will restrict ourselves to perturbing tensors as
$\sum_j\Lambda_{ij} A^{j}$, keeping in mind that the corresponding PEPS is
a ground state of a perturbed Hamiltonian. The fixed points of
$\mathbb{T}$, which are also fixed points of any perturbed transfer matrix
$\mathbb{T}_\Lambda$, correspond to the boundary of those eigenstates of
the parent Hamiltonian, which are also eigenstates of any perturbed parent
Hamiltonian, of the form given in Eq.~\eqref{eq:Hpert}. In other words, fixed points of $\mathbb{T}$ stable under any perturbation arising from $\Lambda$ describe the boundary of ground states. We will refer to them as \textit{stable} fixed points. Concretely, a set of stable fixed points satisfies:
\begin{align}\label{eq:stab}
\mathbb{T}_\infty \mathbb{T}_\Lambda R_i \propto R_i \ \ \forall \ \ \ \Lambda \ ,
\end{align}
where $\mathbb{T}_\infty$ is the projector onto the fixed point space. 

In the following, we will determine the structure of this fixed point
space.  We start in Sec.~\ref{sec:pos} by showing that any set
$\{R_i\}_i$ spanning the fixed point space is stable under perturbations
if the $R_i$ are all positive and mutually orthogonal. We
then continue in Sec.~\ref{sec:fpconst} by showing how to explicitly
construct such a set of stable fixed points for the case where the
transfer operator is unital and has a full rank left positive fixed
points. (Section~\ref{sec:cond}  discusses technical result by
Wolf~\cite{wolf} needed for the proof, showing that the fixed point space
of a unital channel, having a full rank left fixed point, forms an algebra.) Finally, we show in
Sec.~\ref{sec:gen} that a set of stable fixed points can be explicitly
constructed in the same way even if these conditions are not met.

\subsubsection{\label{sec:pos}Conditions for stability}

Let us first consider the case where we are given a basis $\{R_i\}_i$ of
the fixed point space which satisfies that the $R_i$ are all positive 
and moreover mutually orthogonal: $\text{Tr}(R_iR_j) = 0$ for $i\neq j$. We
will show that under these conditions, the fixed point space is stable,
i.e., Eq.~\eqref{eq:stab} holds.

Let $\{L_i\}_i$ be a basis of the fixed point space of the dual transfer operator $\mathbb{T}^*$, which is dual to $\{R_i\}_i$  in the sense that  $\text{Tr}(L_i^\dagger R_j) = \delta_{ij}$. We can use them to write the projector onto the fixed point space $\mathbb{T}_\infty(\rho) = \sum_i R_i\text{Tr}(L_i^\dagger\rho)$ and its dual map $\mathbb{T}^*_\infty(\rho) = \sum_i L_i\text{Tr}(R_i^\dagger \rho)$. Since $\mathbb{T}^*_\infty$ is completely positive we have that $\mathbb{T}^*_\infty(R_j) = \sum_i L_i\text{Tr}(R_i^\dagger R_j) = L_j$ is positive. Here we use positivity and orthogonality of $R_i$. 
In order to prove Eq.~\eqref{eq:stab} it is sufficient to show that $\text{tr}[L_i\mathbb{T}_\Lambda ( R_j)] =0$ if $i\neq j$. Let $M_{ij} = \sum_{\alpha\delta k}\sum_{\beta\gamma}\sqrt{L_i}_{\alpha\beta}B^k_{\beta\gamma}\sqrt{R_j}_{\gamma\delta} \ket{\alpha k\delta}$. Using positivity of both $L_i$ and $R_j$ we obtain the following equation:
\begin{align}\label{eq:pos_orth}
  \text{Tr}[L_i\mathbb{T}_\Lambda ( R_j)] = \braket{M_{ij} |\mathbb{I}\otimes \Lambda \otimes \mathbb{I} | M_{ij}} \ \ \ \forall i,j\ .
\end{align}
Also, by construction of $M_{ij}$ we have that $\langle M_{ij}\vert
M_{ij}\rangle =
\text{tr}[L_i\mathbb{T} ( R_j)] = \delta_{ij}$. Combining these facts we
conclude that indeed that indeed $\text{tr}[L_i\mathbb{T}_\Lambda ( R_j)]
= 0$ if $i\neq j$ and hence that the set of $\{R_i\}_i$ is a stable set of
fixed points.

\subsubsection{\label{sec:cond}Structure of the fixed point space}

Following \cite{wolf}, we will show that the fixed point set of unital
transfer operators, whose dual has a positive full rank fixed point, is an
algebra. The algebra structure will allow us to construct a set of
positive, i.e., stable fixed points in the following subsection. Starting
point is a  Cauchy-Schwarz like inequality for
unital CP maps $\mathbb{T}$~\cite{wolf}:
\begin{align}
\label{eq:cs-type-eq}
\mathbb{T}(AA^\dagger) \geq \mathbb{T}(A)\mathbb{T}(A^\dagger)\ .
\end{align}
This can be verified by taking the Stinespring representation
$\mathbb{T}(A)=V(A\otimes\mathbb{I})V^\dagger$ with $V$ an isometry, i.e.
$V^\dagger V \leq \mathbb{I}$. Let $R$ be a fixed point of the unital CP
map $\mathbb{T}$ and $L$ a fixed point of the dual map $\mathbb{T}^*$ and
consider the equality   
$\text{tr}[L(\mathbb{T}(RR^\dagger) - \mathbb{T}(R)\mathbb{T}(R^\dagger))]
=\text{tr}[\mathbb T^*(L)RR^\dagger - LRR^\dagger]=0$, 
where we have used $\mathbb T(R)=R$ and $\mathbb T^*(L)=L$.
  From Eq.~\eqref{eq:cs-type-eq}
it already follows that $\mathbb{T}(RR^\dagger) -
\mathbb{T}(R)\mathbb{T}(R^\dagger)$ is positive. Hence if $L$ is positive
and has full rank, this equality tells us that
\begin{align}\label{eq:trr}
\mathbb{T}(RR^\dagger) = \mathbb{T}(R)\mathbb{T}(R^\dagger)\ .
\end{align} 
Now, in order to show that the fixed point space of $\mathbb{T}$ is an algebra, take two fixed points $R_1$ and $R_2$ and apply Eq.~\eqref{eq:trr} to $R = R_1+t^*R_2^\dagger$ for some complex $t$:
\begin{align} \nonumber
 &0 = \mathbb{T}(R_1R_1^\dagger) - \mathbb{T}(R_1)\mathbb{T}(R_1^\dagger)\\ \nonumber
 &+t[\mathbb{T}(R_1R_2) - \mathbb{T}(R_1)\mathbb{T}(R_2)]\\ \nonumber
 &+ t^* [\mathbb{T}(R^\dagger_2R_1^\dagger) - \mathbb{T}(R_2^\dagger)\mathbb{T}(R_1^\dagger)]\\
&+|t|^2[\mathbb{T}(R_2^\dagger R_2) - \mathbb{T}(R_2^\dagger)\mathbb{T}(R_2)]\ .
\end{align}
The terms constant and quadratic in $t$ vanish due to Eq.~\eqref{eq:trr}.
By replacing $t\rightarrow it$, adding the two equalities, one ends up
with $\mathbb{T}(R_1R_2) = \mathbb{T}(R_1)\mathbb{T}(R_2)$

\subsubsection{\label{sec:fpconst}Construction of stable fixed points}

Let us now show how the algebra structure can be used to construct
positive orthogonal right fixed points for unital channels.  In the
following subsection, we will then show how to adapt these arguments for
the case of non-unital channels.

In the presence of long range order, there is a fixed point $r_\alpha$ for each symmetry sector $\alpha$. Due to the non-degeneracy of fixed points per symmetry sector we have that ${r}_\alpha{r}_\beta \propto {r}_{\alpha+\beta}$. The algebra of fixed points of $\mathbb{T}$ is hence generated by a single element, in the case of full symmetry breaking being ${r}_1$, which we can choose to normalize to $r_1^N = \mathbb{I}$. We can set all other fixed points such that ${r}_\alpha:= {r}_1^\alpha$. We also have, using non-degeneracy again, that ${r}_1^\dagger = \gamma {r}_1^{-1}$. Taking the $N$-th power of this equation shows that $\gamma$ is an $N$-th root of unity. Multiplying this equation with $r_1$ shows that $\gamma$ should be positive. Hence $\gamma=1$ and  ${r_1}$ and thus also ${r}_\alpha$, is unitary for all $\alpha$. 

The rich structure of the fixed point space of  $\mathbb{ T}$ allows us to define the following orthogonal projectors:
\begin{align}\label{eq:pos_fp}
R_i = \frac{1}{N}\sum_\alpha \omega^{i\alpha} r_\alpha\ .
\end{align}
From ${r}_\alpha{r}_\beta = {r}_{\alpha+\beta}$ it follows that $R_iR_j =
\delta_{ij}R_i$ and from ${r}_\alpha^\dagger = {r}_{N-\alpha}$ it follows
that $R_i^\dagger = R_i$. Hence these fixed points are both orthogonal and
positive.

\subsubsection{\label{sec:gen}Generalization to non-unital transfer
operators}

The transfer operators we are considering are not necessarily unital. The
general idea to fix this is to replace $\mathbb T$ by
$\tilde{\mathbb{T}}:\rho\mapsto
\sqrt{r_0^{-1}}\mathbb{T}\left(\sqrt{r_0}\rho\sqrt{
r_0}\right)\sqrt{r_0^{-1}}$, which is again unital, where $r_0$ is a
positive and full-rank right fixed point. To also cover the general case
where $r_0$ is not full rank, we consider that there exist two maps $V$
and $W$ satisfying $[W^\dagger V,U]= 0$, $VW^\dagger=\mathbb{I}$ and
\begin{align}\label{eq:prop1}
\text{Tr}(Wl^\dagger W^\dagger V rV^\dagger) = \text{Tr}(l^\dagger r) \ ,
\end{align}
for any pair of left and right fixed point $l$ and $r$ of $\mathbb{T}$,
where $Vr_0V^\dagger = \mathbb{I}$ and $Wl_0W^\dagger$ is positive full rank
for some right and left fixed point $r_0$ and $l_0$ with irrep $\alpha=0$,
respectively.  In the Appendix, we show how to explicitly construct such
$V$ and $W$ in the general case; for the generic case where $\mathbb{T}$
already has a full rank positive left and right fixed point, we can choose
$V = \sqrt{r_0}$ and $W=\sqrt{r_0^{-1}}$.

Let us again denote the unique fixed points of $\mathbb T$ in each irrep
sector $\alpha$ by $r_\alpha$ and $l_\alpha$, respectively.  We now
introduce a modified transfer operator $\tilde{\mathbb{T}}=
V\mathbb{T}_\infty(W^\dagger\rho W)V^\dagger$ where again
$\mathbb{T}_\infty$ is the projector onto the fixed point space of
$\mathbb{T}$.  We have that $\tilde{\mathbb{T}}(\rho) =
\sum_i\tilde{r}_\alpha\text{Tr}(\tilde{l}_\alpha^\dagger\rho)$, where
$\tilde{r}_\alpha := Vr_\alpha V^\dagger$ and $\tilde{l}_\alpha :=
Wl_\alpha W^\dagger$. Since $\text{Tr}[\tilde{l}_\alpha^\dagger
\tilde{r}_\beta] = \text{Tr}[{l}_i^\dagger {r}_j]=\delta_{ij}$ (implied by
Eq.~\eqref{eq:prop1}) it follows that the $\tilde{r}_\alpha$ are fixed
points of $\tilde{\mathbb{T}}$, with dual fixed points $\tilde{l}_\alpha$.
Thus, $\tilde{\mathbb{T}}$ is unital (as $\tilde r_0=\mathbb I$) and has a full
rank positive left fixed point (namely $\tilde l_0$). In addition,
$\tilde{\mathbb T}$ has the same $\mathbb Z_N$ symmetry with generator
$\tilde U=VUW^\dagger$, and the $\tilde r_\alpha$ and $\tilde l_\alpha$
transform accordingly. Therefore, we can apply the results of 
Sec.~\ref{sec:fpconst} to find the stable fixed points of $\tilde{\mathbb
T}$, $\tilde R_i = \sum \omega^{i\alpha} \tilde r_\alpha$, as well as
the corresponding
\begin{align}
R_i = \sum \omega^{i\alpha} r_\alpha \ ,
\end{align}
which by construction satisfy $\tilde R_i = V R_i V^\dagger$, and
corresponding left fixed points $L_i=\sum \omega^{i\alpha} l_\alpha$,
$\tilde L_i = W L_i W^\dagger$.  Now consider a perturbation $\mathbb
T_\Lambda$ of $\mathbb T$.  We have that
\[
\mathrm{Tr}[L_i^\dagger\mathbb{T}_\Lambda ( R_j)] 
= 
\mathrm {Tr}[L_i^\dagger\mathbb{T}_\infty \mathbb{T}_\Lambda\mathbb{T}_\infty ( R_j)]
= \mathrm{Tr}[\tilde{L}^\dagger_i\mathbb{\tilde{T}}_\Lambda (
\tilde{R}_j)]\ ,
\]
where we defined $\tilde{\mathbb{T}}_\Lambda(\rho) :=
V\mathbb{T}_\infty\mathbb{T}_\Lambda\mathbb{T}_\infty(W^\dagger\rho
W)V^\dagger$ and used \eqref{eq:prop1}. Since $\tilde{\mathbb
T}=\tilde{\mathbb T}_{\mathbb I}$, $\tilde{\mathbb T}_\Lambda$ is indeed 
a (special) perturbation of $\tilde{\mathbb T}$, and using the result of
Sec.~\ref{sec:pos}, we find that the r.h.s.\ and thus also
$\mathrm{Tr}[L_i^\dagger \mathbb T_\Lambda(R_j)]=0$ unless $i=j$, proving
the stability of the basis $R_i$, $L_j$ against perturbations.

\section{\label{sec:numerics}Numerical study: the 3-state Potts model}

In this section we numerically study symmetry breaking in a quantum model
derived from the classical $3$-state Potts model which has a $\mathbb{Z}_3$
symmetry.  We find that in the ordered phase, corresponding to low
temperatures, there is indeed a degeneracy in the largest eigenvalue of
the transfer operator, which is not the case in the
disordered phase. The corresponding fixed points can be labeled according
to their symmetry. We compare these {symmetric} fixed points to the set of
stable fixed points, which are also fixed points of a perturbed transfer
matrix. First, we check that these are indeed related to each other by a
Fourier transform, see Eq.~\eqref{eq:pos_fp}. And secondly, we compare the
locality of their corresponding boundary Hamiltonians.

\begin{figure}[!t]
  \includegraphics[width=0.3\textwidth]{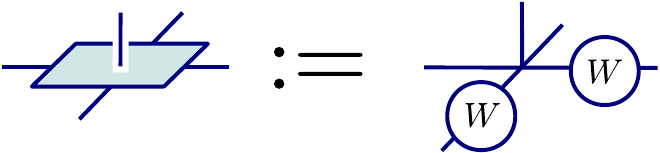}
  \caption{
\label{fig:potts-peps}
The Potts PEPS tensor is defined as a dimension three (bond and physical) delta-tensor contracted with a matrix $M(\beta)$ (see text) on two adjactent virtual indices. }
\end{figure}

The Hamiltonian of the 3 state Potts model is given by $H =
-\sum_{\braket{i,j}}\delta(s_i,s_j)$, where $s_i$ is a classical spin
variable at site $i$ taking values 0, 1 or 2. Its partition function $Z =
\text{Tr} e^{-\tau H}$ at inverse temperature $\tau$ is equal to the norm
squared of a (unnormalized) Rokhsar-Kivelson-type wave function: 
$\ket{\Phi(\tau)}:=\sum_{\{s_i\}} e^{-\tau/2
H(\{s_i\})}\ket{\{s_i\}}$. Indeed, the correlations of $\ket{\Phi(\tau)}$
in the diagonal basis are exactly the same as for the corresponding Potts
model, and it thus exhibits a phase transition at the very same value of
$\tau$. This wave function has an exact PEPS
description~\cite{verstraete:comp-power-of-peps},
cf.~Fig.~\ref{fig:potts-peps}: the PEPS tensors, having
bond dimension $3$, are given by $A^{s_i}_{\alpha\beta\gamma\eta} =
\sum_{\zeta\theta}
\delta_{s_i=\alpha=\beta=\zeta=\theta}W(\tau)_{\zeta\gamma}W(\tau)_{\theta\eta}$.
The matrices $W(\tau)$, sitting on the virtual links, take care of the
contribution $e^{-\tau/2 h(s_i,s_j)}$ of the spins $s_i$ and $s_j$
neighboring the corresponding virtual link. Up to normalization, it is given by
\begin{equation}
W(\tau):=\begin{pmatrix} 1 & a & a\\ a &1 & a\\ a & a & 1\end{pmatrix}\ , 
\end{equation}
where $a=e^{-\tau/2}$. We use the parameter $\theta \in [0,1]$, related
to inverse temperature $\tau$ as  $e^{-\tau/2} = \sin(\pi\theta/2)$ to
interpolate between the ordered and the disorded phases. In terms of this
parameter, the phase transition takes place at $\theta_c \approx 0.4149$.
The matrix $W$ is invariant under a $\mathbb{Z}_3$ action, realized by a
cyclic shift of the basis vectors, hence the full tensor
$A^{s_i}_{\alpha\beta\gamma\delta}$ has a symmetry in the sense of
Eq.~\eqref{symmetry-relation} with 
\begin{equation}
s  = S =\begin{pmatrix} 0 & 1 & 0\\ 0 & 0 & 1\\ 1 & 0 & 0 \end{pmatrix}\ .
\end{equation}\\
\subsubsection{Phase transition}
In figure Fig.~\ref{potts-figures}(a) we report on the spectral gap between the largest and second largest eigenvalue of the transfer operator. These values are obtained by exact diagonalization of a transfer operators of a different sizes $N_v$ (6, 8 or 10). We clearly observe a degeneracy in the ordered phase and a non-zero gap in the disorderd phase. The transition becomes sharper as we increase $N_v$, ie. when finite size effects are decreased.

\begin{figure}[t]
\includegraphics[width=0.9\columnwidth]{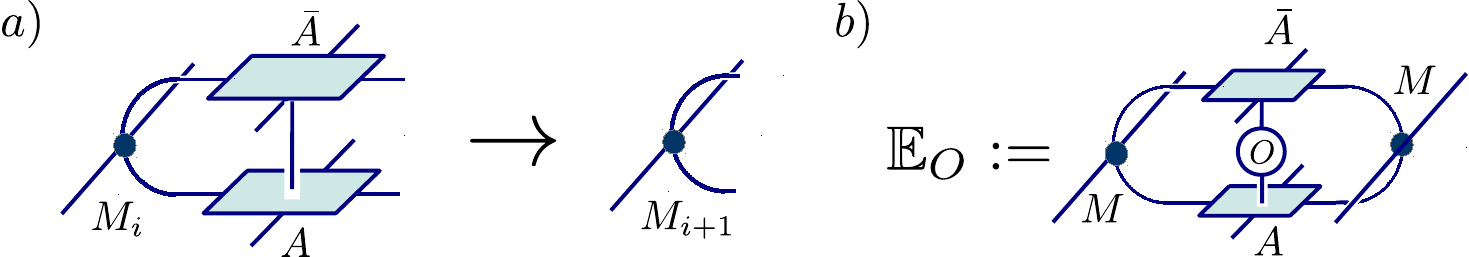}
\caption{\label{fig:iMPS}
iMPS method, cf.~text. (a) Iteration step $\mathbb TM_i\to M_{i+1}$. (b)
The operator $\mathbb E_O$ used to compute correlation length and order
parameter.}
\end{figure}

Additionally, to verify the critical value $\theta_c $, we use an infinite
Matrix Product State (``iMPS'') algorithm to determine correlation lengths
and magnetization.  The iMPS algorithm produces a translational invariant
MPS description of the fixed point of a transfer operator for which $N_v
\rightarrow\infty$, thereby avoiding any finite size effects. Inspired by
the fact that $\mathbb{T}^\infty$ projects onto the fixed point space, such
algorithms start off with an arbitrary translational invariant MPS $M$,
from which a sequence of MPSs are created recursively by application of
the transfer operator: $\mathbb{T}M_{i}\rightarrow M_{i+1}$. Crucially, at
each step truncation takes place to keep the bond dimension from growing.
This is done by truncating the singular values of $\sqrt{L}\sqrt{R}$,
where $L$ and $R$ are the left and right largest eigenvector of the
transfer matrix $\mathbb{E}$ of the MPS.

The fixed points of the transfer operator can be used to efficiently
contract a tensor network representing the expectation values of local
operators. As a result, such values can be written purely in terms of the
MPS tensor $M$ of the fixed point. Let $\mathbb{E}_O = \sum_i M^i \otimes
B_O\otimes\bar{M^i}$ be the dressed transfer matrix corresponding to the
fixed point MPS \footnote{When left and right fixed points are described
by different MPS $M_L$ and $M_R$ one should consider the mixed transfer
matrix $\mathbb{E}_O = \sum_i M_L^i \otimes B_O\otimes \bar{M_R^i}$}.
The correlation length is
related to the spectral gap of $\mathbb{E}_\mathbb{I}$ as $\xi = 
-(\log\frac{\lambda_1}{\lambda_0})^{-1}$, where $\lambda_i$ is the $i$-th
eigenvector of $\mathbb{E}_\mathbb{I}$. The
expectation value of a local operator is given by
$1/\lambda_0\braket{l|\mathbb{E}_O|r}$, where $r$ and $l$ are the right
and left eigenvectors of $\mathbb{E}_\mathbb{I}$, respectively. We
calculate  the correlation length, Fig.~\ref{potts-figures}(b), and the
the expectation value of $Z = \diag{1, & \omega, & \bar\omega}$, which
acts as an order parameter since it does not commute with the symmetry
$u$, Fig.~\ref{potts-figures}(c).  The data shows a diverging correlation
length, together with the onset of the order parameter, at the expected
value of transition. Moreover, we take confidence from here, that a
moderate bond dimension is sufficient to describe the fixed point, away
from the critical point, since the correlation length saturates with
increasing $\chi$.

Note that from the onset of an order parameter we can conclude that we
have obtained a symmetry broken fixed point from the iMPS algorithm, even
though we have not explicitly broken the symmetry of the transfer
operator. Indeed, it is well-known that variational methods tend to
spontaneously break symmetries since they generally prefer states with
fewer long-range correlations. It is however not clear a priori whether
this symmetry breaking is the same as the one derived in 
the Section \ref{sec:proof} using \emph{physical} perturbations of the
transfer operator, although it is plausible to assume that the numerical
inaccuracies giving rise to the observed symmetry breaking act in an
equivalent way.

\subsubsection{Stability of fixed points}
We continue by explicitly verifying the relation between {stable} fixed
points and {symmetric} fixed points in the ordered phase. For finite
$N_v$, exact diagonalization easily allows to extract the symmetric fixed
points by using projectors on the different symmetry sectors. Note that
the symmetric fixed points are defined up to a phase. We set this phase by
insisting that the largest eigenvalues of the fixed points be positive.
The stable fixed points are obtained by explicitly breaking the symmetry
of the PEPS tensor (and hence of the transfer operator) by $A\rightarrow
(\mathbb{I}+\epsilon Z) A$ where again $Z = \diag{1, & \omega, &
\bar\omega}$.

\begin{figure}[t]
  \includegraphics[width=0.5\textwidth]{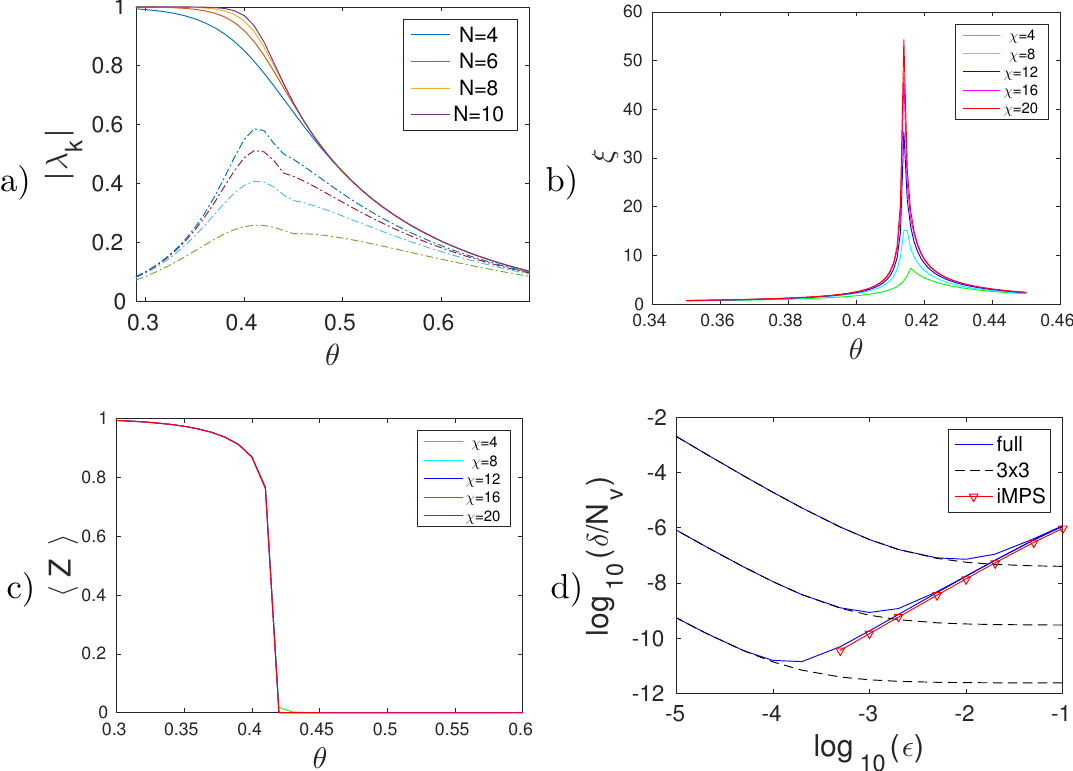}
  \caption{\label{potts-figures}3-state Potts PEPS model a) Second- and
fourth-largest eigenvalue of the transfer operator for finite size (second
and third are (numerically) exactly degenerate) b) correlation length of
the fixed point iMPS for increasing iMPS bond dimension $\chi$  c) order
parameter $\langle Z \rangle$ measured on iMPS fixed point d) perturbation
strength against infidelity per site for perturbation scaling analysis.
blue: exact diagonalization on $N_v=10$ (compressed), red: iMPS, dashed:
exact diagonalization ($N_v=10$) on $\{r_0,r_\omega,r_{\bar\omega}\}$
subspace ($\theta=0.25$)} 
\end{figure}

In Fig.~\ref{potts-figures}(d) we report on the infidelity per site
$\delta =  \min_i(1-{\braket{R_i|R_\epsilon}})/N_v$ between the stable
fixed points $R_\epsilon$ and the Fourier transform of the symmetric fixed
points $R_i$ given by Eq.~\eqref{eq:sym_fp} at $\theta = 0.25 < \theta_c$
(well into the ordered phase). We observe infidelities of the order
of $10^{-5}  - 10^{-10}$ and thus conclude that the stable fixed points
are indeed well approximated by the Fourier transform of symmetric fixed
points. We further observe that the infidelity per site scales as
$\epsilon^2$ and is independent of system size $N_v$. This can be
understood from perturbation theory. Up to quadratic terms in $\epsilon$, the
perturbed transfer operator is given by $\mathbb{T} + \epsilon
\sum_{k=1}^{N_v} \mathbb{T}_{Z+Z^\dagger}^{[k]}$. The perturbed fixed
point is given by $R_\epsilon \approx
\frac{1}{\sqrt{1+\epsilon^2\,\mathrm{Tr}({R_i^\perp}^2)}}(R_i + \epsilon R_i^\perp)$
where $R_i^\perp$ is orthogonal to $R_i$. This explains the $\epsilon$
dependency of the infidelity. Moreover $R_i^\perp =
\sum_{k=1}^{N_v}\sum_n\frac{V^{[k]}_{n0}}{1 - \lambda_n} \rho_n$ where
$V^{[k]}_{n0}$ is the coupling between the fixed point $R_i$ and
eigenstate $\rho_n$ (with eigenvalue $\lambda_n<1$) due to the
perturbation $\mathbb{T}_{Z+Z^\dagger}^{[k]}$. The fact that
$|R_i^\perp|^2$ scales linearly with $N_v$ is consistent with $R_i$ being
finitely correlated: $V^{[k]}_{n0}$ only couples to states  $\rho_n$ which
differ from $R_i$ in a neighborhood of site $k$.

The above analysis breaks down as soon as the perturbation $\epsilon$ is
weaker than the coupling between the stable fixed points due to finite
size effects. This explains the increase in infidelity below a certain
$\epsilon_t$. As expected, the value of $\epsilon_t$ decreases with
increasing size.  To further analyze this effect, we have also considered
the restriction of the perturbed transfer operator $\mathbb{T}_\epsilon$
to the the 3 dimensional fixed point space of $\mathbb{T}$, i.e. the space
spanned by $r_\alpha$ \footnote{Strictly speaking, due to finite size
effects, these are not all fixed points}. Finding the fixed points of
$\Pi\mathbb{T}_\epsilon \Pi$ (where $\Pi$ is the orthogonal projection onto the
fixed point space of $\mathbb{T}$) amounts to simply diagonalizing a 3 by
3 matrix of the form $T(\epsilon) = \mathbb{I}+\Delta+\epsilon P$. Here
$\Delta$ is the finite size effect and is hence diagonal in the symmetric
fixed point basis, and $P$ is the perturbation.  Again we report on
infidelity $\delta =  \min_i(1-{\braket{R_i|R^r_\epsilon}})/N_v$ with
$R^r_\epsilon$ the fixed point of $\Pi\mathbb{T}_\epsilon\Pi$ and $R_i$ again
the Fourier transform of the symmetric fixed point of the unperturbed
transfer operator,  see Fig.~\ref{potts-figures}(d). For small $\epsilon$
this infidelity is completely identical to earlier obtained infidelities
at same system sizes. Thus the competition between splitting due to finite
size effects and perturbation completely explains the observed
infidelities. Interestingly, for large $\epsilon$ the fidelity saturates.
In this regime, eigenstates of $T(\epsilon)$ are simply eigenstates of $P$
but are apparently not exactly equal to the Fourier transform of the
symmetric fixed points, as we would have expected from the discussion in
Section \ref{sec:proof}. Note that, in proving that the stable fixed
points are of the form $R_i = \sum_\alpha \omega^{i\alpha}r_\alpha$ we
assumed an exact degeneracy, any finite size splitting could hence also
alter the relation between the stable and the symmetric fixed points.

To avoid finite size effects, we also calculate the fixed point of the
transfer operator with $N_v\rightarrow\infty$ using the iMPS method. In
this case, we do not verify Eq.~\eqref{eq:pos_fp} explicitly but rather
verify that the obtained fixed point has large overlap with the stable
fixed point. Overlaps between MPSs $\ket{\phi_M}$ and $\ket{\phi_N}$ are
given by $\left(\frac{\lambda_{MN}^2}{\lambda_{MM}\lambda_{NN}}\right)^N$
where $\lambda_{MN}$ is the largest eigenvalue of the mixed transfer
matrix $\mathbb{E} = \sum_i M^i \otimes \bar{N^i}$. Hence the infidelity
per site, when it is small, is well approximated by $\delta = 1 -
\frac{\lambda_{MN}^2}{\lambda_{MM}\lambda_{NN}}$. We find that the
infidelity is again of the order of $10^{-5} - 10^{-10}$, see
Fig.~\ref{potts-figures}(d). Moreover, we find that in it scales as
$\epsilon^2$ also for small $\epsilon$.

\subsubsection{Locality of fixed points}

To further quantify the difference between stable and symmetric fixed
point, we study the locality of their corresponding boundary Hamiltonian.
The entanglement or boundary Hamiltonian is defined via its Gibbs state
\begin{equation}
\rho_A = \exp(-H_E) \Leftrightarrow H_E = -\log(\rho_A) \ ,
\end{equation}
where $\rho_A$ is a fixed point of the transfer operator. In the ordered
phases it is a priori unclear which of the degenerate fixed points one
should consider although Hermiticity of the Hamiltonian does imply
positivity of the corresponding fixed point. Note that all stable fixed
points and fixed points in the trivial symmetry sectors are positive. We
report on the locality of the boundary Hamiltonian by decomposing it into
$k$-local terms: 
\begin{equation}
H = \sum_{i,k} h_{i}^{(k)} \Gamma_i^{(k)}, \quad \Gamma_i^{(k)} = \mathbb{I} \otimes \cdots \mathbb{I}\otimes \underbrace{X \otimes \cdots Y}_{k \text{ sites apart}}\otimes\mathbb{I}\otimes \cdots \mathbb{I}\ .
\end{equation}
The terms $\Gamma_i^{(k)}$ have non-trivial support on only $k$ consecutive sites. Specifically, we construct $\Gamma_i^{(k)}$ by taking tensor products of Gell-Mann-matrices which constitute an orthonormal basis for $3\times3$ Hermitian matrices (under the Hilbert-Schmidt inner product). Let $w_k = \sum_i (h_i^{(k)})^2$ be the total  strength of all $k$-local contributions. We report on $w_k$ as a function of $k$ for boundary Hamiltonians obtained at different temperatures. The fixed points used here are obtained by first using the iMPS algorithm to generate the MPS representation of a fixed point. We then use $8$ copies of these matrices to approximate the fixed point of a transfer operator of size $N_v=8$. This method allows use to study a finite size transfer operator whilst minimizing finite size effects.

\begin{figure}[t]
  \includegraphics[width=\columnwidth]{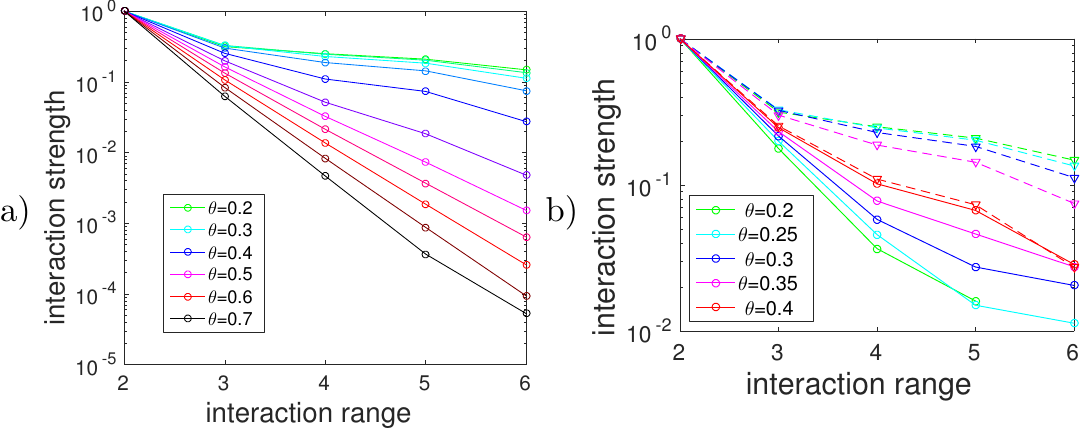}
  \caption{Interaction strength ($\log_{10}$) vs. interaction range for the 3-state Potts model PEPS. Computed with iMPS and put on a cylinder of length 8. Fig. a) The weights of the symmetric fixed point clearly become non-local in the symmetry broken phase (since we get the symmetry broken fixed points with iMPS, we restore the symmetry with a twirl) b) Now in the symmetry broken phase $\theta < 0.41$ we compare the weights of the boundary Hamiltonians obtained from symmetric (dashed lines) and symmetry broken fixed points (solid lines), note that the color ordering matches, i.e. while the symmetry broken one gets more local, the symmetric one gets increasingly non-local.}
  \label{potts_model_imps_boundary_hamiltonian}
\end{figure}

In the disordered phase, the $w_k$ decrease exponentially with $k$,
indicating a \textit{local} boundary Hamiltonian, see
Fig.~\ref{potts_model_imps_boundary_hamiltonian}(a). The exponent
increases as one approaches the phase transition. At the phase transition the
weights $w_k$ still seem to decay exponentially with $k$. Beyond the phase
transition, we study both the locality of the fixed point obtained from
iMPS (i.e.\ the stable fixed point) as well as a symmetrized version $r_0 =
\sum_i U^iRU^{i\dagger}$. In
Fig.~\ref{potts_model_imps_boundary_hamiltonian}(b) it can be seen that as
we go away from phase transition, the locality of the symmetric fixed
point increases whereas that of the stable fixed point decreases. Hence we
find that the boundary Hamiltonian corresponding to the symmetric fixed
point is less local than the boundary Hamiltonians corresponding to
the stable fixed points, reinforcing the perspective that the symmetry
broken fixed points are the ones which are physically relevant.

\section{Conclusions}

In this work, we have studied the occurence of symmetry breaking and long-range order in
PEPS models with a $\mathbb Z_N$ symmetry.  Firstly, we have shown that long-range
order is accompanied by a degeneracy in the spectrum of the transfer
operator, with the gap closing at least as $1/N_v$. We have subsequently
studied how the PEPS reacts to physical perturbations, i.e., those
corresponding to perturbations of the Hamiltonian, and have determined the
fixed points of the transfer operator (i.e., the boundary conditions to
the system) which are stable under perturbations and thus describe the
symmetry broken states.  We have found that these states are uniquely
determined by the symmetry structure of the transfer operator together
with the requirement that they are positive, and are given by the Fourier
transform of the fixed points in the individual symmetry sectors.  We have
finally numerically studied symmetry breaking in a PEPS model related to
the $\mathbb Z_3$ Potts model, where we confirmed the relation between
long-range order and the degeneracy of the transfer operator, as well as
the form of symmetry broken fixed points.  We subsequently computed the
entanglement Hamiltonian both for the symmetric and the symmetry broken
fixed points, and found that unlike the symmetric one, the symmetry broken
fixed points give rise to a quasi-local entanglement Hamiltonian,
demonstrating the local nature of the entanglement Hamiltonian also for
symmetry broken phases.

\begin{acknowledgments}

We acknowledge helpful discussions with Mohsin Iqbal. This work has been
supported by the Alexander von Humboldt foundation, the European Union
through the ERC grant WASCOSYS (No.\ 636201), and by JARA-HPC through
grants jara0092 and jara0111.  
\end{acknowledgments}

\bibliography{ref}

\begin{thebibliography}{18}%
\makeatletter
\providecommand \@ifxundefined [1]{%
 \@ifx{#1\undefined}
}%
\providecommand \@ifnum [1]{%
 \ifnum #1\expandafter \@firstoftwo
 \else \expandafter \@secondoftwo
 \fi
}%
\providecommand \@ifx [1]{%
 \ifx #1\expandafter \@firstoftwo
 \else \expandafter \@secondoftwo
 \fi
}%
\providecommand \natexlab [1]{#1}%
\providecommand \enquote  [1]{``#1''}%
\providecommand \bibnamefont  [1]{#1}%
\providecommand \bibfnamefont [1]{#1}%
\providecommand \citenamefont [1]{#1}%
\providecommand \href@noop [0]{\@secondoftwo}%
\providecommand \href [0]{\begingroup \@sanitize@url \@href}%
\providecommand \@href[1]{\@@startlink{#1}\@@href}%
\providecommand \@@href[1]{\endgroup#1\@@endlink}%
\providecommand \@sanitize@url [0]{\catcode `\\12\catcode `\$12\catcode
  `\&12\catcode `\#12\catcode `\^12\catcode `\_12\catcode `\%12\relax}%
\providecommand \@@startlink[1]{}%
\providecommand \@@endlink[0]{}%
\providecommand \url  [0]{\begingroup\@sanitize@url \@url }%
\providecommand \@url [1]{\endgroup\@href {#1}{\urlprefix }}%
\providecommand \urlprefix  [0]{URL }%
\providecommand \Eprint [0]{\href }%
\providecommand \doibase [0]{http://dx.doi.org/}%
\providecommand \selectlanguage [0]{\@gobble}%
\providecommand \bibinfo  [0]{\@secondoftwo}%
\providecommand \bibfield  [0]{\@secondoftwo}%
\providecommand \translation [1]{[#1]}%
\providecommand \BibitemOpen [0]{}%
\providecommand \bibitemStop [0]{}%
\providecommand \bibitemNoStop [0]{.\EOS\space}%
\providecommand \EOS [0]{\spacefactor3000\relax}%
\providecommand \BibitemShut  [1]{\csname bibitem#1\endcsname}%
\let\auto@bib@innerbib\@empty
\bibitem [{\citenamefont {Griffiths}(1966)}]{Griffiths}%
  \BibitemOpen
  \bibfield  {author} {\bibinfo {author} {\bibfnamefont {R.~B.}\ \bibnamefont
  {Griffiths}},\ }\href {\doibase 10.1103/PhysRev.152.240} {\bibfield
  {journal} {\bibinfo  {journal} {Phys. Rev.}\ }\textbf {\bibinfo {volume}
  {152}},\ \bibinfo {pages} {240} (\bibinfo {year} {1966})}\BibitemShut
  {NoStop}%
\bibitem [{\citenamefont {Koma}\ and\ \citenamefont
  {Tasaki}(1993)}]{komotasaki}%
  \BibitemOpen
  \bibfield  {author} {\bibinfo {author} {\bibfnamefont {T.}~\bibnamefont
  {Koma}}\ and\ \bibinfo {author} {\bibfnamefont {H.}~\bibnamefont {Tasaki}},\
  }\href {http://projecteuclid.org/euclid.cmp/1104254136} {\bibfield  {journal}
  {\bibinfo  {journal} {Comm. Math. Phys.}\ }\textbf {\bibinfo {volume}
  {158}},\ \bibinfo {pages} {191} (\bibinfo {year} {1993})}\BibitemShut
  {NoStop}%
\bibitem [{\citenamefont {m.~b. hastings}(2006)}]{hastings:locally}%
  \BibitemOpen
  \bibfield  {author} {\bibinfo {author} {\bibnamefont {m.~b. hastings}},\
  }\href@noop {} {\bibfield  {journal} {\bibinfo  {journal} {phys. rev. b}\
  }\textbf {\bibinfo {volume} {73}},\ \bibinfo {pages} {085115} (\bibinfo
  {year} {2006})},\ \Eprint {http://arxiv.org/abs/cond-mat/0508554}
  {cond-mat/0508554} \BibitemShut {NoStop}%
\bibitem [{\citenamefont {Molnar}\ \emph {et~al.}(2015)\citenamefont {Molnar},
  \citenamefont {Schuch}, \citenamefont {Verstraete},\ and\ \citenamefont
  {Cirac}}]{molnar:thermal-peps}%
  \BibitemOpen
  \bibfield  {author} {\bibinfo {author} {\bibfnamefont {A.}~\bibnamefont
  {Molnar}}, \bibinfo {author} {\bibfnamefont {N.}~\bibnamefont {Schuch}},
  \bibinfo {author} {\bibfnamefont {F.}~\bibnamefont {Verstraete}}, \ and\
  \bibinfo {author} {\bibfnamefont {J.~I.}\ \bibnamefont {Cirac}},\ }\href@noop
  {} {\bibfield  {journal} {\bibinfo  {journal} {Phys. Rev. B}\ }\textbf
  {\bibinfo {volume} {91}},\ \bibinfo {pages} {045138} (\bibinfo {year}
  {2015})},\ \Eprint {http://arxiv.org/abs/arXiv:1406.2973} {arXiv:1406.2973}
  \BibitemShut {NoStop}%
\bibitem [{\citenamefont {Verstraete}\ and\ \citenamefont
  {Cirac}(2004)}]{verstraete:2D-dmrg}%
  \BibitemOpen
  \bibfield  {author} {\bibinfo {author} {\bibfnamefont {F.}~\bibnamefont
  {Verstraete}}\ and\ \bibinfo {author} {\bibfnamefont {J.~I.}\ \bibnamefont
  {Cirac}},\ }\href@noop {} {\  (\bibinfo {year} {2004})},\ \Eprint
  {http://arxiv.org/abs/cond-mat/0407066} {cond-mat/0407066} \BibitemShut
  {NoStop}%
\bibitem [{\citenamefont {Orus}(2014)}]{orus:tn-review}%
  \BibitemOpen
  \bibfield  {author} {\bibinfo {author} {\bibfnamefont {R.}~\bibnamefont
  {Orus}},\ }\href@noop {} {\bibfield  {journal} {\bibinfo  {journal} {Ann.
  Phys.}\ }\textbf {\bibinfo {volume} {349}},\ \bibinfo {pages} {117} (\bibinfo
  {year} {2014})},\ \Eprint {http://arxiv.org/abs/arXiv:1306.2164}
  {arXiv:1306.2164} \BibitemShut {NoStop}%
\bibitem [{\citenamefont {Perez-Garcia}\ \emph {et~al.}(2008)\citenamefont
  {Perez-Garcia}, \citenamefont {Verstraete}, \citenamefont {Cirac},\ and\
  \citenamefont {Wolf}}]{ParentHamiltonian}%
  \BibitemOpen
  \bibfield  {author} {\bibinfo {author} {\bibfnamefont {D.}~\bibnamefont
  {Perez-Garcia}}, \bibinfo {author} {\bibfnamefont {F.}~\bibnamefont
  {Verstraete}}, \bibinfo {author} {\bibfnamefont {J.~I.}\ \bibnamefont
  {Cirac}}, \ and\ \bibinfo {author} {\bibfnamefont {M.~M.}\ \bibnamefont
  {Wolf}},\ }\href@noop {} {\bibfield  {journal} {\bibinfo  {journal} {Quantum
  Inf. Comput.}\ }\textbf {\bibinfo {volume} {8}},\ \bibinfo {pages} {0650}
  (\bibinfo {year} {2008})},\ \Eprint {http://arxiv.org/abs/arXiv:0707.2260}
  {arXiv:0707.2260} \BibitemShut {NoStop}%
\bibitem [{\citenamefont {Cirac}\ \emph {et~al.}(2011)\citenamefont {Cirac},
  \citenamefont {Poilblanc}, \citenamefont {Schuch},\ and\ \citenamefont
  {Verstraete}}]{cirac:peps-boundaries}%
  \BibitemOpen
  \bibfield  {author} {\bibinfo {author} {\bibfnamefont {J.~I.}\ \bibnamefont
  {Cirac}}, \bibinfo {author} {\bibfnamefont {D.}~\bibnamefont {Poilblanc}},
  \bibinfo {author} {\bibfnamefont {N.}~\bibnamefont {Schuch}}, \ and\ \bibinfo
  {author} {\bibfnamefont {F.}~\bibnamefont {Verstraete}},\ }\href@noop {}
  {\bibfield  {journal} {\bibinfo  {journal} {Phys. Rev. B}\ }\textbf {\bibinfo
  {volume} {83}},\ \bibinfo {pages} {245134} (\bibinfo {year} {2011})},\
  \Eprint {http://arxiv.org/abs/arXiv:1103.3427} {arXiv:1103.3427} \BibitemShut
  {NoStop}%
\bibitem [{\citenamefont {Yang}\ \emph {et~al.}(2014)\citenamefont {Yang},
  \citenamefont {Lehman}, \citenamefont {Poilblanc}, \citenamefont {Acoleyen},
  \citenamefont {Verstraete}, \citenamefont {Cirac},\ and\ \citenamefont
  {Schuch}}]{yang:peps-edgetheories}%
  \BibitemOpen
  \bibfield  {author} {\bibinfo {author} {\bibfnamefont {S.}~\bibnamefont
  {Yang}}, \bibinfo {author} {\bibfnamefont {L.}~\bibnamefont {Lehman}},
  \bibinfo {author} {\bibfnamefont {D.}~\bibnamefont {Poilblanc}}, \bibinfo
  {author} {\bibfnamefont {K.~V.}\ \bibnamefont {Acoleyen}}, \bibinfo {author}
  {\bibfnamefont {F.}~\bibnamefont {Verstraete}}, \bibinfo {author}
  {\bibfnamefont {J.}~\bibnamefont {Cirac}}, \ and\ \bibinfo {author}
  {\bibfnamefont {N.}~\bibnamefont {Schuch}},\ }\href@noop {} {\bibfield
  {journal} {\bibinfo  {journal} {Phys. Rev. Lett.}\ }\textbf {\bibinfo
  {volume} {112}},\ \bibinfo {pages} {036402} (\bibinfo {year} {2014})},\
  \Eprint {http://arxiv.org/abs/arXiv:1309.4596} {arXiv:1309.4596} \BibitemShut
  {NoStop}%
\bibitem [{\citenamefont {Verstraete}\ \emph {et~al.}(2006)\citenamefont
  {Verstraete}, \citenamefont {Wolf}, \citenamefont {Perez-Garcia},\ and\
  \citenamefont {Cirac}}]{verstraete:comp-power-of-peps}%
  \BibitemOpen
  \bibfield  {author} {\bibinfo {author} {\bibfnamefont {F.}~\bibnamefont
  {Verstraete}}, \bibinfo {author} {\bibfnamefont {M.~M.}\ \bibnamefont
  {Wolf}}, \bibinfo {author} {\bibfnamefont {D.}~\bibnamefont {Perez-Garcia}},
  \ and\ \bibinfo {author} {\bibfnamefont {J.~I.}\ \bibnamefont {Cirac}},\
  }\href@noop {} {\bibfield  {journal} {\bibinfo  {journal} {Phys.\ Rev.\
  Lett.}\ }\textbf {\bibinfo {volume} {96}},\ \bibinfo {pages} {220601}
  (\bibinfo {year} {2006})},\ \Eprint {http://arxiv.org/abs/quant-ph/0601075}
  {quant-ph/0601075} \BibitemShut {NoStop}%
\bibitem [{\citenamefont {Schuch}\ \emph {et~al.}(2013)\citenamefont {Schuch},
  \citenamefont {Poilblanc}, \citenamefont {Cirac},\ and\ \citenamefont
  {Perez-Garcia}}]{schuch:topo-top}%
  \BibitemOpen
  \bibfield  {author} {\bibinfo {author} {\bibfnamefont {N.}~\bibnamefont
  {Schuch}}, \bibinfo {author} {\bibfnamefont {D.}~\bibnamefont {Poilblanc}},
  \bibinfo {author} {\bibfnamefont {J.~I.}\ \bibnamefont {Cirac}}, \ and\
  \bibinfo {author} {\bibfnamefont {D.}~\bibnamefont {Perez-Garcia}},\
  }\href@noop {} {\bibfield  {journal} {\bibinfo  {journal} {Phys. Rev. Lett.}\
  }\textbf {\bibinfo {volume} {111}},\ \bibinfo {pages} {090501} (\bibinfo
  {year} {2013})},\ \Eprint {http://arxiv.org/abs/arXiv:1210.5601}
  {arXiv:1210.5601} \BibitemShut {NoStop}%
\bibitem [{\citenamefont {{Rispler}}\ \emph {et~al.}(2015)\citenamefont
  {{Rispler}}, \citenamefont {{Duivenvoorden}},\ and\ \citenamefont
  {{Schuch}}}]{rispler:lro-and-ssb-in-peps}%
  \BibitemOpen
  \bibfield  {author} {\bibinfo {author} {\bibfnamefont {M.}~\bibnamefont
  {{Rispler}}}, \bibinfo {author} {\bibfnamefont {K.}~\bibnamefont
  {{Duivenvoorden}}}, \ and\ \bibinfo {author} {\bibfnamefont {N.}~\bibnamefont
  {{Schuch}}},\ }\href {\doibase 10.1103/PhysRevB.92.155133} {\bibfield
  {journal} {\bibinfo  {journal} {\prb}\ }\textbf {\bibinfo {volume} {92}},\
  \bibinfo {eid} {155133} (\bibinfo {year} {2015})},\ \Eprint
  {http://arxiv.org/abs/1505.04217} {arXiv:1505.04217 [cond-mat.str-el]}
  \BibitemShut {NoStop}%
\bibitem [{\citenamefont {Kaplan}\ \emph {et~al.}(1989)\citenamefont {Kaplan},
  \citenamefont {Horsch},\ and\ \citenamefont {von~der Linden}}]{KHvdL}%
  \BibitemOpen
  \bibfield  {author} {\bibinfo {author} {\bibfnamefont {T.}~\bibnamefont
  {Kaplan}}, \bibinfo {author} {\bibfnamefont {P.}~\bibnamefont {Horsch}}, \
  and\ \bibinfo {author} {\bibfnamefont {W.}~\bibnamefont {von~der Linden}},\
  }\href {\doibase 10.1143/JPSJ.58.3894} {\bibfield  {journal} {\bibinfo
  {journal} {Journal of the Physical Society of Japan}\ }\textbf {\bibinfo
  {volume} {58}},\ \bibinfo {pages} {3894} (\bibinfo {year} {1989})},\ \Eprint
  {http://arxiv.org/abs/http://dx.doi.org/10.1143/JPSJ.58.3894}
  {http://dx.doi.org/10.1143/JPSJ.58.3894} \BibitemShut {NoStop}%
\bibitem [{Note1()}]{Note1}%
  \BibitemOpen
  \bibinfo {note} {Let $\tau _{N_v} = \protect \qopname \relax m{lim}_{N_h} N_h
  \sigma ^2_{N_h,N_v}$, and let $S:=\protect \qopname \relax m{lim}_{N_v}\tau
  _{N_v}$. Then, for any $\epsilon >0$, \begin {align*} &\exists N_v^0 \
  \forall N_v\ge N_v^0\protect \tmspace +\thinmuskip {.1667em}:\ {\setbox \z@
  \hbox {\frozen@everymath \@emptytoks \mathsurround \z@ $\nulldelimiterspace
  \z@ \left |\vcenter to1.5\big@size {}\right .$}\box \z@ } \protect \genfrac
  {}{}{}1{1}{N_v} \tau _{N_v} - S {\setbox \z@ \hbox {\frozen@everymath
  \@emptytoks \mathsurround \z@ $\nulldelimiterspace \z@ \left |\vcenter
  to1.5\big@size {}\right .$}\box \z@ } < \protect \frac {\epsilon }{2} \unhbox
  \voidb@x \hbox {\hskip 1em\relax and} \\ &\exists N_h^0(N_v) \ \forall N_h\ge
  N_h^0(N_v):\protect \tmspace +\medmuskip {.2222em} \protect \frac
  {1}{N_v}{\setbox \z@ \hbox {\frozen@everymath \@emptytoks \mathsurround \z@
  $\nulldelimiterspace \z@ \left |\vcenter to1.5\big@size {}\right .$}\box \z@
  } N_h \sigma _{N_h,N_v}^2 - \tau _{N_v} {\setbox \z@ \hbox {\frozen@everymath
  \@emptytoks \mathsurround \z@ $\nulldelimiterspace \z@ \left |\vcenter
  to1.5\big@size {}\right .$}\box \z@ } < \protect \frac {\epsilon }{2}\ .
  \protect \hspace *{-2em} \end {align*} Thus, \protect \[ \left |\protect
  \frac {N_h}{N_v}\sigma _{N_h,N_v}^2-S\right |\le \epsilon \ , \protect \] and
  finally \protect \[ \sigma ^2=\protect \qopname \relax m{lim}_{N_h,N_v\to
  \infty } \sigma _{N_h,N_v}^2 \le \protect \frac {N_h}{N_v}\sigma
  _{N_h,N_v}^2=S \ , \protect \] as long as we couple the limits such that both
  $N_h\ge N_h^0(N_v)$ and $N_h\ge N_v$. (If $S=\infty $, the inequality
  \protect \textup {\hbox {\mathsurround \z@ \protect \normalfont
  (\ignorespaces \ref {eq:onelimit-twolimits}\unskip \@@italiccorr )}} holds
  trivially.) Let us note that for normal $\protect \mathbb T$, the convergence
  in Eq.~\protect \textup {\hbox {\mathsurround \z@ \protect \normalfont
  (\ignorespaces \ref {ourresult}\unskip \@@italiccorr )}}, Ref.~\cite
  {rispler:lro-and-ssb-in-peps}, yields a scaling $N_h^0(N_v)\propto
  N_v/(1-|\lambda _\alpha (N_v)|)$, such that a non-zero $\sigma ^2$ for all
  isotropically coupled limits $N_h/N_v=\protect \mathrm {const.}$ is
  sufficient to infer that $|\lambda _\alpha |\rightarrow 1$ as $N_v\to \infty
  $.}\BibitemShut {Stop}%
\bibitem [{\citenamefont {Fannes}\ \emph {et~al.}(1992)\citenamefont {Fannes},
  \citenamefont {Nachtergaele},\ and\ \citenamefont {Werner}}]{fannes1992}%
  \BibitemOpen
  \bibfield  {author} {\bibinfo {author} {\bibfnamefont {M.}~\bibnamefont
  {Fannes}}, \bibinfo {author} {\bibfnamefont {B.}~\bibnamefont
  {Nachtergaele}}, \ and\ \bibinfo {author} {\bibfnamefont {R.~F.}\
  \bibnamefont {Werner}},\ }\href
  {http://projecteuclid.org/euclid.cmp/1104249404} {\bibfield  {journal}
  {\bibinfo  {journal} {Comm. Math. Phys.}\ }\textbf {\bibinfo {volume}
  {144}},\ \bibinfo {pages} {443} (\bibinfo {year} {1992})}\BibitemShut
  {NoStop}%
\bibitem [{\citenamefont {{Wolf}}(2012)}]{wolf}%
  \BibitemOpen
  \bibfield  {author} {\bibinfo {author} {\bibfnamefont {M.~M.}\ \bibnamefont
  {{Wolf}}},\ }\href@noop {} {\enquote {\bibinfo {title} {{Quantum Channels and
  Operations, Guided Tour}},}\ } (\bibinfo {year} {2012})\BibitemShut {NoStop}%
\bibitem [{Note2()}]{Note2}%
  \BibitemOpen
  \bibinfo {note} {When left and right fixed points are described by different
  MPS $M_L$ and $M_R$ one should consider the mixed transfer matrix $\protect
  \mathbb {E}_O = \DOTSB \sum@ \slimits@ _i M_L^i \otimes B_O\otimes \protect
  \mathaccentV {bar}016{M_R^i}$}\BibitemShut {NoStop}%
\bibitem [{Note3()}]{Note3}%
  \BibitemOpen
  \bibinfo {note} {Strictly speaking, due to finite size effects, these are not
  all fixed points}\BibitemShut {NoStop}%
\end{thebibliography}%

\appendix*

\section{Construction of $V$ and $W$}

Consider a transfer operator (CP-map) $\mathbb{T}$ with symmetry $U$ as in
Eq.~\eqref{eq:symT}. In this Appendix we will construct the $V$ and $W$ used
in Sect.~\ref{sec:gen}, i.e., which satisfy that there exist a left and right fixed point
$R_0$ and $L_0$ of $\mathbb{T}$ such that
$VR_0V^\dagger=\mathbb{I}$ and $WL_0W^\dagger$ is positive and full rank,
$VW^\dagger = \mathbb{I}$, $[W^\dagger V,U] =0$, and
Eq.~\eqref{eq:prop1} holds for any pair of fixed points $R$ and $L$ of
$\mathbb{T}$. 

Let $R_0 = \mathbb{T}_\infty(\mathbb{I})$ and $L_0 = \mathbb{T}^*_\infty(\mathbb{I})$. Both $R_0$ and $L_0$ are positive due to positivity of $\mathbb{T}$ and are symmetric due to Eq.~\eqref{eq:symT}. Let $P_1$ be the isometry ($P_1P^\dagger_1=\mathbb{I}$) such that $P^\dagger_1P_1$ projects onto the support of $R_0$ and let $P_2$ be isometry ($P_2P^\dagger_2=\mathbb{I}$) such that $P^\dagger_2P_2$ projects onto the support of $\hat{L}_0 = \sqrt{\hat{R}_0}P_1 L_0P^\dagger_1\sqrt{\hat{R}_0}$, with $\hat R_0 := P_1 R_0 P^\dagger_1$. The maps $V$ and $W$ are given by:
\begin{align}
 V = P_2 \sqrt{\hat{R}_0^{-1}} P_1 \ ,\\
 W = P_2 \sqrt{\hat{R}_0} P_1 \ .
\end{align}
Note that $\hat{R}_0$ is by construction invertible. It can be
straightforwardly checked that $VR_0V^\dagger=\mathbb{I}$. By definition
of $P_2$, $WL_0W^\dagger$ is full rank. It is also positive since $L_0$ is
positive. Also it is obvious that $VW^\dagger = \mathbb{I}$. It remains to check that $[W^\dagger V,U] =0$ and Eq.~\eqref{eq:prop1} holds for any pair of fixed points $R$ and $L$ of $\mathbb{T}$.

The symmetry condition follows from the symmetry of $R_0$ and $L_0$. From $[R_0,U] =0$ it follows that $[P^\dagger_1P_1,U] =0$. This assures that $\hat{U} := P_1 UP^\dagger_1$ is unitary. It commutes with $\hat R_0$ and hence also with  $\sqrt{\hat{R}_0}$ and  $\sqrt{\hat{R}_0^{-1}}$. From $[R_0,U] =0$ and the previous facts, it follows that $\hat{U}$ commutes with $\hat L_0$. Using the same argument it follows that $[P^\dagger_2P_2,\hat{U}] =0$ showing that $U' = P_2\hat{U}P^\dagger_2$ is unitary. Combining everything it follows that $U'V = VU$ and $U'W = WU$ showing that $[W^\dagger V,U] =0$.

In order to show that Eq.~\eqref{eq:prop1} holds for any pair of fixed points $R$ and $L$ of $\mathbb{T}$ we will first show that the support and range of any fixed point $R$ of $\mathbb{T}$ is contained in the support of $R_0$. Let $R$ be some Hermitian fixed point, and let $\epsilon>0$ be sufficiently small such that $\mathbb{I} \pm \epsilon R$ is positive definite. The fixed point $\mathbb{T}^\infty(\mathds{1}\pm\epsilon X)=  R_0 \pm\epsilon R$ has to be positive by definition of CP, but the r.h.s. fails to be positive as soon as $R$ has support outside $R_0$. Since the fixed point space is closed under Hermitian conjugation (both $R + R^\dagger$ and $i(R - R^\dagger)$ are also fixed points for any non Hermitian $R$) the support and range of any fixed point $R$ is contained in the support of $R_0$.

From the restriction of the support and range of fixed points it follows that
\begin{align}\nonumber
 \text{Tr}(L^\dagger R)  =& \text{Tr}(L^\dagger P_1^\dagger P_1 R P_1^\dagger P_1) \\\nonumber
 =& \text{Tr}( \sqrt{\hat{R_0}} P_1L^\dagger P_1^\dagger \sqrt{\hat{R}_0} \sqrt{\hat{R}_0^{-1}} P_1 R P_1^\dagger \sqrt{\hat{R}_0^{-1}}) \\\nonumber
 =& \text{Tr}( P_2^\dagger P_2\sqrt{ \hat{R}_0} P_1L^\dagger P_1^\dagger \sqrt{\hat{R}_0} P_2^\dagger P_2 \cdot \\ \nonumber
 &\quad\sqrt{\hat{R}_0^{-1}} P_1 R P_1^\dagger \sqrt{\hat{R}_0^{-1}} ) \\
 =& \text{Tr}( W L^\dagger W^\dagger V RV^\dagger ) \ .
\end{align}
The third equality follows from the fact that the support and range of any fixed point $L$ of $\mathbb{T}^*$ is contained in the support of $L_0$ and hence that the support and range of $\sqrt{ \hat{R}_0} P_1L P_1^\dagger \sqrt{\hat{R}_0}$ is contained in the support of $\hat L_0$. In more detail let $v\in\text{kern}(\hat L_0) \Rightarrow v\hat L_0 v = 0$ from which it follows (due to positivity of $L_0$) that $P_1^\dagger \sqrt{\hat{R}_0} v \in \text{kern}(L_0)\subset \text{kern}(L)$. Thus the support of $\sqrt{ \hat{R}_0} P_1L P_1^\dagger \sqrt{\hat{R}_0}$ is contained in the support of $\hat L_0$. The restriction on its range follows from the invariance of the fixed point space under Hermitian conjugation.

\end{document}